\documentclass[prl,aps,twocolumn,showpacs,superscriptaddress,nofootinbib]{revtex4}
\usepackage{graphicx} % Include figure files
\usepackage{dcolumn}  % Align table columns on decimal point
\usepackage{bm}       % bold math
\usepackage{amsmath}
\usepackage{epsfig}

\def\rr{\textbf{r}} \def\pp{\textbf{p}} \def\EE{\textbf{E}} 
\def\xx{\hat{\bf x}}  \def\yy{\hat{\bf y}}  \def\zz{\hat{\bf z}}
\def\ssplus{\hat{\bm{\sigma}}^{+}}
\def\ssminus{\hat{\bm{\sigma}}^{-}}

\def\aeff{\bar{\alpha}_{\rm{eff}}}

\def\GG{\overline{G}}
\def\aa{\overline{\alpha}}
\def\Im{\mbox{Im}}

\begin{document}
\title{Lateral Casimir force on a rotating particle near a planar surface}
\author{Alejandro Manjavacas}
\email[Corresponding author: ]{manjavacas@unm.edu}
\affiliation{Department of Physics and Astronomy, University of New Mexico, Albuquerque, New Mexico 87131, United States}
\author{Francisco J. Rodr\'{i}guez-Fortu\~{n}o}
\email[Corresponding author: ]{francisco.rodriguez_fortuno@kcl.ac.uk}
\affiliation{Department of Physics, King's College London, London WC2R 2LS, UK}
\author{F. Javier Garc\'{\i}a de Abajo}
\affiliation{ICFO - Institut de Ciencies Fotoniques, The Barcelona Institute of Science \& Technology, Castelldefels (Barcelona), Spain}
\affiliation{ICREA - Instituci\'o Catalana de Recerca i Estudis Avan\c{c}ats (ICREA), Barcelona, Spain}
\author{Anatoly V. Zayats}
\affiliation{Department of Physics, King's College London, London WC2R 2LS, UK}

%\date{\today}

\begin{abstract}
We study the lateral Casimir force experienced by a particle that rotates near a planar surface. The origin of this force lies in the symmetry breaking induced by the particle rotation in the vacuum and thermal fluctuations of its dipole moment, and, therefore, in contrast to lateral Casimir forces previously described in the literature for corrugated surfaces, it exists despite the translational invariance of the planar surface. 
Working within the framework of fluctuational electrodynamics, we derive analytical expressions for the lateral force and analyze its dependence on the geometrical and material properties of the system. In particular, we show that the direction of the force can be controlled by adjusting the particle-surface distance, which may be exploited as a new mechanism to manipulate nanoscale objects.
\end{abstract}
\pacs{42.50.Wk, 42.50.Lc, 45.20.dc, 78.70.-g}
\maketitle

% --- intro ---

Fluctuation-induced forces exist between polarizable atoms, nonpolar molecules, and structured materials, emerging as a result of vacuum and thermal fluctuations that involve virtual electromagnetic excitations.
Generally known as van der Waals or London dispersion forces at short range distances, Casimir-Polder forces when taking retardation into account, and Casimir-Lifshitz forces when including material dispersion, these are generally referred to as Casimir forces \cite{C1948,L07,DMR11}. There is strong evidence that various phenomena in nature such as adhesion, friction, wetting, and stiction are a result of these forces \cite{MC10}, and therefore, their study can shed light into the mechanical behavior of nanodevices, where these forces may play a dominant role \cite{B07_2,RCJ11}. 

Casimir forces are typically attractive, acting along symmetry directions (\textit{e.g.}, the normal to the interacting surfaces \cite{HJM02,SKD11}). However, if the surfaces are corrugated, these forces may have a component parallel to the surface, which is commonly referred to as \textit{lateral} \cite{GK1997,EHG01,CMK02_2,DML08}. The lateral Casimir force has been successfully measured and has been argued to enable interesting applications such as contactless transmission of lateral motion \cite{EHG03,BKM04,AMG07,CKM10,NMM10}. Nevertheless, the force still acts along local surface-normal directions and arises due to the broken mirror symmetry introduced by the corrugations. Therefore, it strongly depends on the mutual geometric lateral displacement between the corrugations on both surfaces, becoming zero when they are aligned and the lateral mirror symmetry is recovered. An alternative symmetry breaking leading to lateral forces may rely on rotational motion of the involved bodies.

\begin{figure}
\begin{center}
\includegraphics[width=70mm,angle=0,clip]{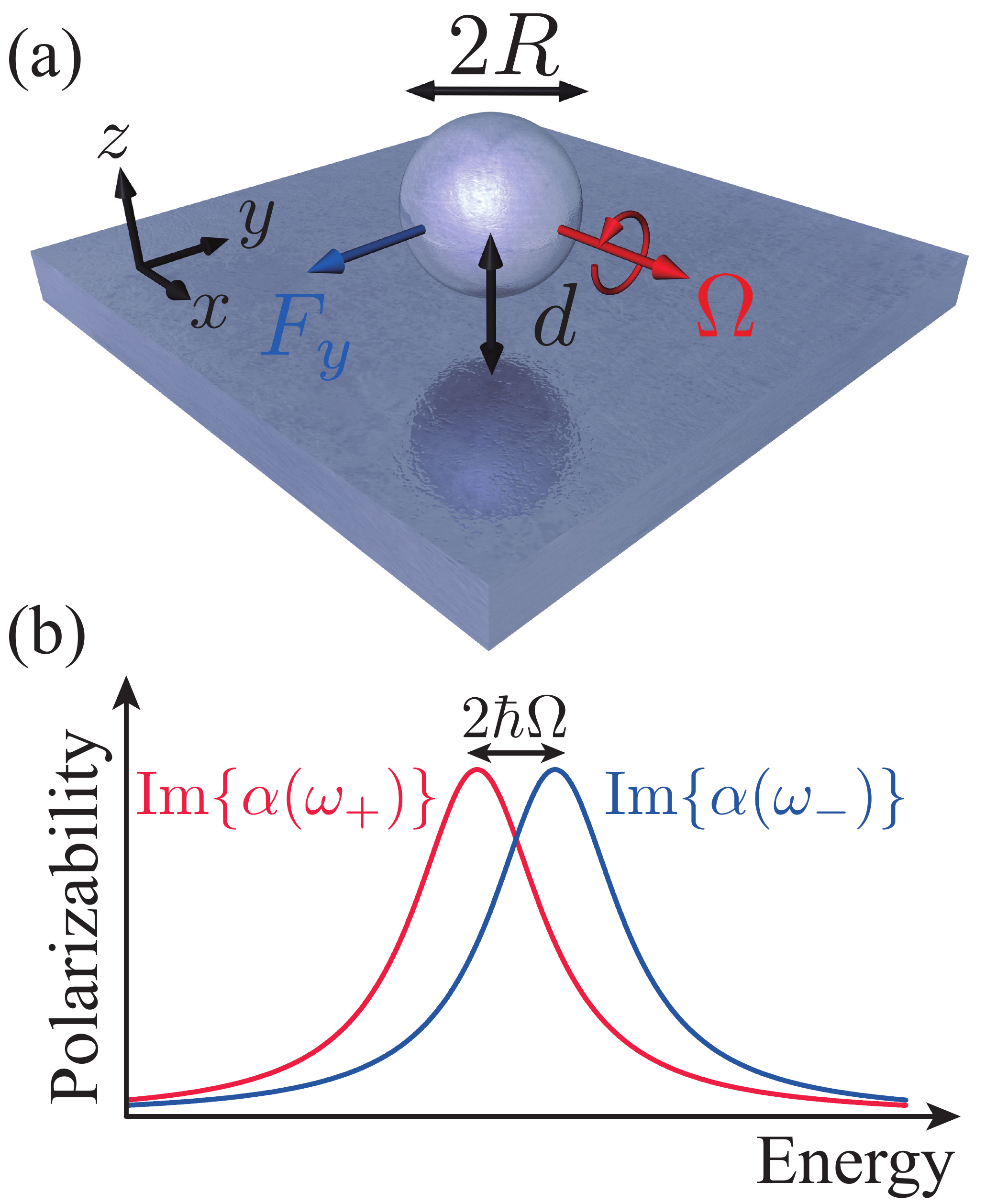}
\caption{(a) Sketch of the system under study. A spherical nanoparticle of radius $R$ is rotating with angular frequency $\Omega$ at a distance $d$ above a planar surface. Due to the rotation, the particle experiences a lateral force $F_y$. (b) Schematic explaining the origin of the lateral force: the resonances of the imaginary part of the left- and right-handed components of the particle polarizability are spectrally separated by a distance equal to twice the rotation energy $2\hbar\Omega$.} \label{fig1}
\end{center}
\end{figure}

In this Letter, we describe a lateral Casimir force that acts on a rotating particle near an ideally flat surface. This lateral force is directed parallel to the surface. The geometry under consideration [Fig.~\ref{fig1}(a)] has translational symmetry in the direction of the force, however, the symmetry is broken by the rotation of the particle, leading to the observed force. 
Rotating particles have been shown to experience Casimir frictional torques that slow down their motion \cite{ama7,ama9,ama19}. Here, we analyze a qualitatively different effect: a lateral force that pushes the particle parallel to the surface and whose direction and magnitude is determined by the sense and frequency of rotation, the particle-surface distance, and the materials from which the particle and the surface are made. 
The force does not depend on the lateral position of the particle due to the translational invariance of the surface and is 
consistent with the frictional force predicted to exist between surfaces in relative uniform motion \cite{P97}.
%, thus making it fundamentally different from previously studied lateral Casimir forces.
The system and the force investigated here constitute a Casimir analogue of a mechanical wheel rotating and moving over a planar surface, but with no contact required. The origin of this force can be traced back to the recently discovered spin-direction locking of electromagnetic evanescent waves \cite{BN12,BBN14,ABN15}, an example of spin-orbit coupling of light \cite{BRN15}. Lateral optical forces \cite{SBC15,SKN15,REM15} naturally arise from an asymmetric scattering by circularly polarized dipoles into electromagnetic modes of any neighboring surface or waveguide \cite{RMG13,MC13,LR14}. In a similar way, the rotating particle experiences an imbalance between left- and right-handed helicities in the vacuum and thermal fluctuations associated with its electromagnetic response, which ultimately leads to the lateral Casimir force predicted in this work.

\begin{figure}
\begin{center}
\includegraphics[width=70mm,angle=0,clip]{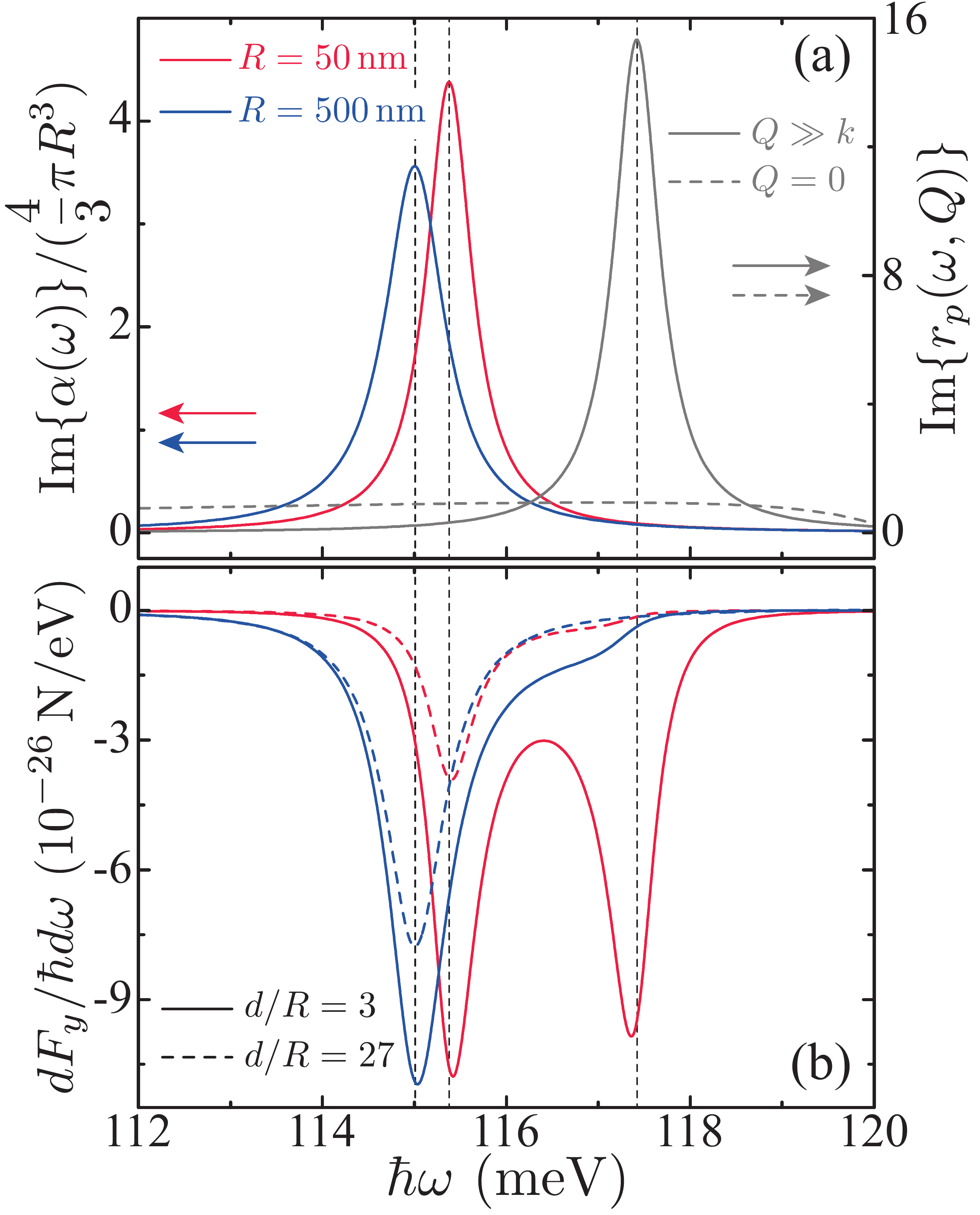}
\caption{(a) Imaginary part of the polarizability of spherical SiC nanoparticles (left scale) with radius $R=50\,$nm (red curve) and $R=500\,$nm (blue curve), and imaginary part of the reflection coefficient a SiC surface (right scale) calculated for $Q=0$ (gray dashed curve) and for $Q\gg k$ (gray solid curve). (b) Frequency integrand of the lateral force [Eq.~(\ref{lateralforcef})] corresponding to the two nanoparticles of panel (a) when they are placed at a distance $d$ from a SiC surface. The curves for $d/R=27$ are multiplied by a factor $200$ to improve visibility. Moreover, we choose a rotation frequency $\Omega/2\pi=1\,$kHz and surface and particle temperatures $T_0=T_1=300\,$K.} \label{fig2}
\end{center}
\end{figure}

{\it Theoretical model.--} We consider a spherical particle small enough to be adequately described within the dipolar limit through a frequency-dependent polarizability $\alpha(\omega)$. The particle rotates around the $x$-axis, parallel to a planar surface located at the $z=0$ plane, with rotation frequency $\Omega$, as shown in Fig.~\ref{fig1}(a).
The rotation modifies the interaction of the particle with the vacuum and thermal electromagnetic fields. In particular, in a frame rotating with the particle, an external circularly polarized (CP) electromagnetic field is 
perceived with a reduced ($\omega_-=\omega-\Omega$) or increased ($\omega_+=\omega+\Omega$) frequency depending on the sense of rotation of the particle relative to the handedness of the field. This results in the splitting of the polarizability of the non-rotating particle into two components $\alpha_{+}(\omega)=\alpha(\omega_-)$ and $\alpha_{-}(\omega)=\alpha(\omega_+)$ associated with the unit vectors of opposite CP field helicities $\ssplus = (1/\sqrt{2})(\yy+i\zz)$ and $\ssminus = (1/\sqrt{2})(\yy-i\zz)$  [Fig.~\ref{fig1}(b)]. In the basis defined by $\{\xx, \ssplus, \ssminus\}$, the effective particle polarizability reduces to
\begin{equation}
\aeff(\omega)
= \begin{pmatrix}
\alpha(\omega) & 0 & 0\\
0 & \alpha(\omega_-) & 0\\
0 & 0 & \alpha(\omega_+)
\end{pmatrix}.\nonumber
\end{equation}
Because the vacuum and thermal fluctuations of the dipole moment of the particle are determined by its effective polarizability through the fluctuation-dissipation theorem (FDT) \cite{N1928,CW1951,NH06,ama7}, the imbalance in the circularly polarized components results in asymmetric fluctuations that are the origin of the lateral force, as discussed above \cite{RMG13,MC13,LR14,SBC15,SKN15,REM15}.

%Furthermore, a generalization of this expression for anisotropic particles is derived in the supplementary material [XX]. 
%  A dipole with circularly polarized components spinning around an axis $\hat{\bf{s}}$ near a surface normal to $\zz$ will experience a lateral optical force \cite{SBC15,SKN15,REM15} in the direction $\hat{\bf{s}} \times \zz$.

\begin{figure*}
\begin{center}
\includegraphics[width=170mm,angle=0,clip]{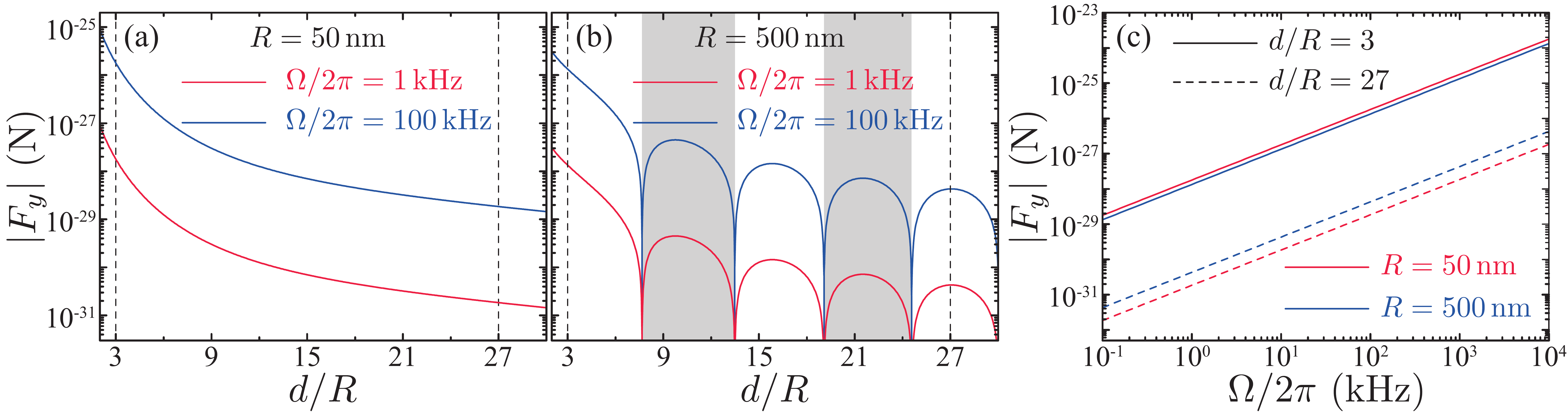}
\caption{(a,b) Lateral Casimir force experienced by a SiC spherical nanoparticle with $R=50\,$nm (a) and $R=500\,$nm (b) rotating near a SiC surface, plotted as a function of the particle-surface distance, $d$, for two different rotation frequencies: $\Omega/2\pi=1\,$kHz (red curves) and  $\Omega/2\pi=100\,$kHz (blue curves). Shaded areas indicate the distances for which the lateral force is positive. (c) Lateral Casimir force for the same nanoparticles as in panels (a,b) plotted as a function of the rotation frequency, $\Omega$, for particle-surface distances as indicated in the legends, which correspond to the dashed vertical lines in panels (a,b). The surface and particle temperatures are $T_0=T_1=300\,$K.} \label{fig3}
\end{center}
\end{figure*}

In order to obtain an analytical expression for the lateral Casimir force, we start by considering the electromagnetic force acting on a dipole $\mathbf{p}$ in an electric field $\mathbf{E}$, which can be written as $\mathbf{F} = \sum_{i}p_i \nabla E_i$  \cite{GA1980}, with $i=x,y,z$. The vacuum and thermal fluctuations causing the Casimir force come from two different sources: (i) fluctuations of the dipole moment of the particle $\mathbf{p}^{\rm{fl}}$, and (ii) fluctuations of the field $\mathbf{E}^{\rm{fl}}$ generated by current fluctuations in the surface. As these two sources of fluctuations originate in different systems, they are uncorrelated, and therefore, we can write the lateral Casimir force as  $F_y= \sum_i\langle p_i^{\rm fl}\partial_y E_i^{\rm ind}+p_i^{\rm ind} \partial_y E_i^{\rm fl}\rangle$, where $\langle \rangle$ stands for the average over fluctuations, which we perform using the FDT. Now, expressing the induced field in terms of the fluctuating dipole with the help of the surface Green function $\overline{G}$, and the induced dipole in terms of the fluctuating field as $\mathbf{p}^{\rm{ind}} =\aeff\mathbf{E}^{\rm{fl}}$, we obtain (see Supplemental Material \cite{EPAPS} for details)
\begin{align}\label{lateralforcef}
F_y=\frac{\hbar}{\pi} &\int_{0}^\infty d\omega \mbox{Im} \left\{\partial_y G_{yz}(\omega)-\partial_y G_{zy}(\omega)\right\}\nonumber \\
&\times \left[\mbox{Im} \left\{ \alpha(\omega_+) \right\}  N(\omega_+)-\mbox{Im} \left\{ \alpha(\omega_-) \right\}  N(\omega_-)\right],
\end{align} 
where $N(\omega_{\pm})=n(T_1,\omega_{\pm})-n(T_0,\omega)$, with $n(T_i,\omega) = [\exp(\frac{\hbar\omega}{k_{ B} T_i}) - 1]^{-1}$ being the Bose-Einstein distribution at temperature $T_i$, and  $T_0$ and $T_1$ the temperatures of the surface and the particle, respectively. Interestingly, the lateral Casimir force is finite even for $T_1=T_0=0\,$K, and, as expected, it vanishes as $\Omega \to 0$. The gradient of the surface Green function can be calculated as (see Supplemental Material \cite{EPAPS} for details)
\begin{equation}\label{gradientGlateral}
\partial_y G_{yz} =  -\partial_y G_{zy}=\frac1{2}\int_0^\infty dQ  e^{2 i k_z d} Q^3  r_p(\omega,Q),
\end{equation}
where the integral is performed over the transverse wave vector $Q$, $k_z = (k^2-Q^2)^{1/2}$ is the wave vector along $z$, $k = \omega/c$ is the free-space wave number, $d$ is the particle-surface distance, and $r_p(\omega,Q)$ is the Fresnel reflection coefficient of the surface for $p$-polarized waves.
It is important to remark that we only consider the surface component of the Green function because the free-space part does not contribute to the lateral force \cite{RCJ11}.
One should also notice that Eq.~(\ref{gradientGlateral}) appears in the calculation of the lateral force acting on a circularly polarized dipole oscillating at a frequency $\omega$ above a surface \cite{REM15}. This implies that the lateral Casimir force can be recast as the frequency integral of the dipole lateral force weighted by the appropriate frequency-dependent fluctuation terms in Eq.~(\ref{lateralforcef}). 

{\it Numerical results.--} Using Eqs.~(\ref{lateralforcef}) and (\ref{gradientGlateral}), we can numerically compute the lateral Casimir force for different scenarios. The material composition and size of the particle determine its isotropic non-rotating polarizability $\alpha(\omega)$ that appears in Eq.~(\ref{lateralforcef}), while the reflection coefficient $r_p(\omega,Q)$, and therefore $\partial_y G_{yz}$,  is controlled by the surface properties [see Eq.~(\ref{gradientGlateral})]. 
Here we choose silicon carbide (SiC) for both the particle and the substrate. SiC is a polaritonic material that supports phonon polaritons and its dielectric function can be modeled as $\varepsilon(\omega)=\varepsilon_{\infty}\left[1-(\omega^2_{\rm L}-\omega^2_{\rm T})/(\omega^2_{\rm T}-\omega^2-i\omega\gamma)\right]$, where $\varepsilon_{\infty}=6.7$, $\hbar\omega_{\rm T}=98.3\,$meV,  $\hbar\omega_{\rm L}=120\,$meV, and $\hbar\gamma=0.59\,$meV \cite{P1985}.

We gain insight into the behavior of the lateral force by examining the integrand of Eq.~(\ref{lateralforcef}), $dF_y/d\omega$. To that end, we plot in Fig.~\ref{fig2} (a) the imaginary part of the polarizability of two SiC particles (left scale) with radius $R=50\,$nm (red curve) and $R=500\,$nm (blue curve), as obtained from the dipolar Mie coefficient \cite{paper112}, which therefore includes retardation that permits us to extend our results up to $kR\sim1$. Indeed, retardation is already visible for the $R=500\,$nm particle polarizability, resulting in resonance redshift and broadening with respect to that of the $R=50\,$nm particle [\textit{cf.} red and blue curves in panel (a)].  Our theoretical model is based on the dipolar approximation and therefore we expect it to be inaccurate for particle-surface distances for which the dipolar plasmon of the particle is modified by hybridization with higher order modes. Panel (a) also shows the imaginary part of the reflection coefficient of a SiC surface (right scale) in the limits $Q=0$ (gray dashed curve) and $Q\gg k$ (gray solid curve).
The integrand of Eq.~(\ref{lateralforcef}) is plotted in Fig.~\ref{fig2}(b) for the same particles and surface as in panel (a), assuming a rotation frequency $\Omega/2\pi=1\,$kHz, temperatures $T_0=T_1=300\,$K, and particle-surface distances $d=3R$ (solid curves) and $d=27R$ (dashed curves). The integrand, which can be interpreted as a force spectral density, is proportional to the difference between ${\rm Im}\{\alpha(\omega_+)\}N(\omega_+)$ and ${\rm Im}\{\alpha(\omega_-)\}N(\omega_-)$. 
It should be noted that, under realistic conditions, the rotation frequency is much smaller than the particle resonance frequency ($\omega_0\sim100\,$THz).
This, allows us to approximate ${\rm Im}\{\alpha(\omega_+)\}N(\omega_+)-{\rm Im}\{\alpha(\omega_-)\} N(\omega_-) \approx 2 \Omega{\rm Im}\{\alpha(\omega)\} \partial n(T,\omega)/\partial \omega$, retaining only linear terms in $\Omega$, and assuming $T_0=T_1=T$. Therefore, the spectral force density, at first order in $\Omega$, is determined by the product of ${\rm Im}\{\alpha(\omega)\}$ and ${\rm Im}\{\partial_yG_{yz}\}$, which in turn is controlled by ${\rm Im}\{r_p\}$.
This is clearly seen in Fig.~\ref{fig2}(b) for the $R=50\,$nm particle when $d=3R$ (red solid curve), for which $dF_y/d\omega$ displays two peaks corresponding to ${\rm Im}\{\alpha\}$ and ${\rm Im}\{r_p\}$ for $Q\gg k$ (non-retarded limit), respectively. 
The second peak becomes less visible for $R=500\,$nm (blue solid curve), and completely disappears for larger separations (dashed curves) due to retardation effects, as expected from the non-resonant behavior of ${\rm Im}\{r_p\}$ when $Q=0$.

Figure \ref{fig3} shows calculations of the lateral Casimir force as a function of particle-surface separation for SiC particles of radius $R=50\,$nm (a) and $R=500\,$nm (b), rotating near a SiC surface at $1\,$kHz (red curves) and $100\,$kHz (blue curves). As expected, the lateral force shows a decreasing trend with distance. Interestingly, this trend is accompanied by an oscillatory behavior (shaded areas correspond to positive forces), which arises from the exponential term of Eq.~(\ref{gradientGlateral}), with a period of $\sim\lambda_0/2$, where $\lambda_0=2\pi c /\omega_0\approx 5.4\,\mu$m is the particle resonance wavelength. 
The sign oscillation of the lateral Casimir force implies that the direction and magnitude of the force can be controlled or even suppressed by choosing the appropriate particle-surface separation.
The dependence of the lateral Casimir force on the rotation frequency is examined in Fig.~\ref{fig3}(c) under the same conditions as in panels (a,b). As anticipated from the analysis of the force spectral density, the value of $|F_y|$ increases linearly with $\Omega$ within the range of rotation frequencies under consideration, for which $\Omega \ll \omega_0$. It is important to remark that $\Omega$ can approach $\omega_0$ in systems consisting of materials with resonances at low-phonon frequencies, for which the dependence on $\Omega$ can be more complex.

\begin{figure}
\begin{center}
\includegraphics[width=70mm,angle=0,clip]{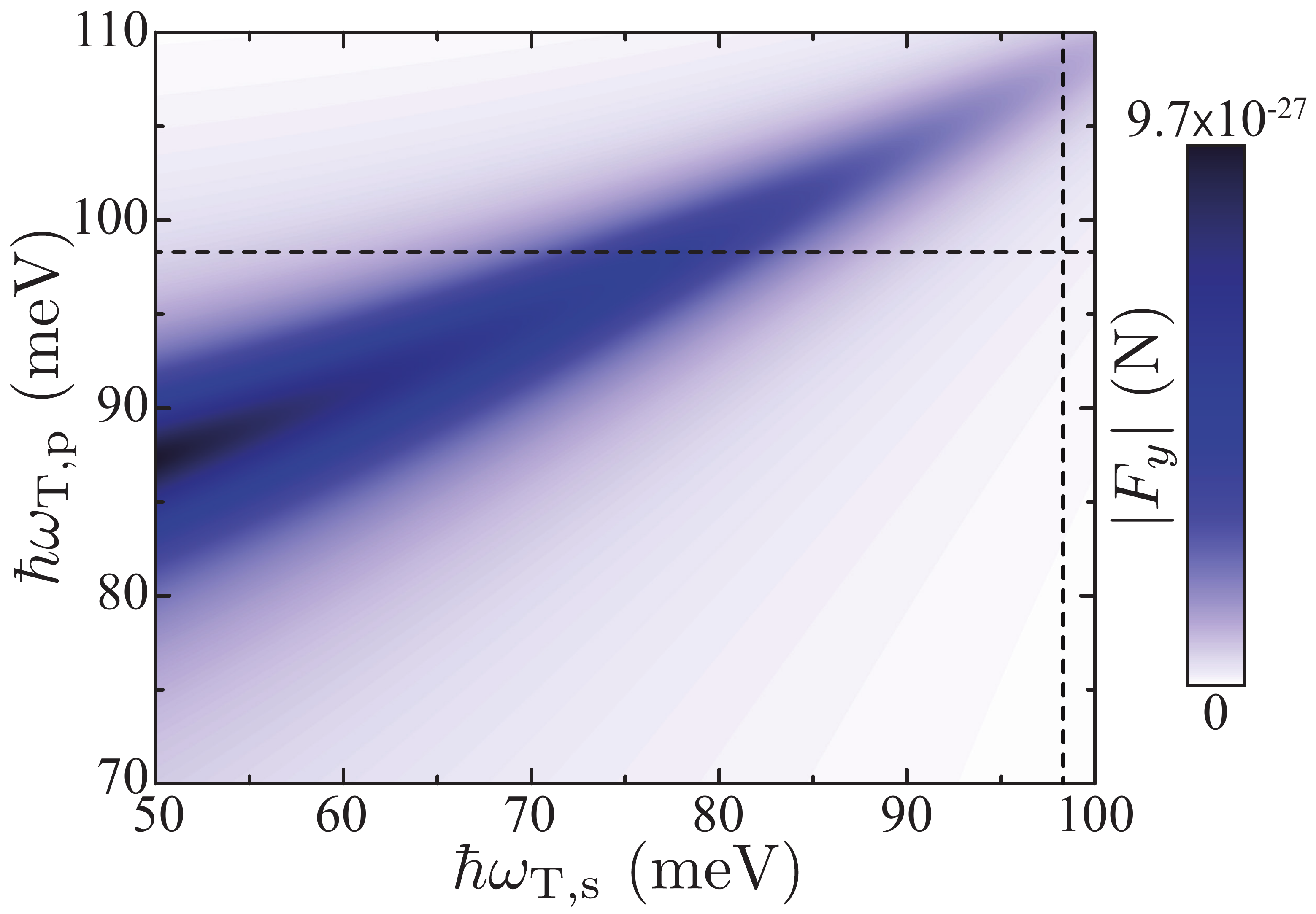}
\caption{Lateral force experienced by a spherical nanoparticle of radius $R=50\,$nm plotted as a function of the transversal phonon polariton energy of the particle $\hbar\omega_{\rm T,p}$, and the surface $\hbar\omega_{\rm T,s}$. Other parameters of the particle and surface dielectric function are those of SiC (see text), the particle-surface distance is $d=3R$, the rotation frequency is $\Omega/2\pi=1\,$kHz, and the temperatures are $T_0=T_1=300\,$K. The dashed lines correspond to $\hbar\omega_{\rm T}=98.3\,$eV as in SiC.} \label{fig4}
\end{center}
\end{figure}

The results in Fig.~\ref{fig3} show that the most advantageous situation to achieve large lateral Casimir forces involves particles rotating at high frequencies, placed close to the surface. In such cases, as discussed above, the force spectral density displays two different peaks; one associated with the polarizability of the particle and the other with the reflection coefficient of the surface [see red solid curve in Fig.~\ref{fig2}(b)]. Therefore, a way to enhance the force would consist in bringing together these two resonances by using particles and surfaces made of different materials. 
This possibility is explored in Fig.~\ref{fig4}, where the lateral Casimir force is plotted varying $\omega_{\rm T}$ for the particle and the surface materials: $\omega_{\rm T,p}$ and $\omega_{\rm T,s}$, respectively. 
For simplicity, we keep the values of $\varepsilon$, $\omega_{\rm L}$, and $\gamma$ the same as in SiC. Examining Fig.~\ref{fig4}, we observe that the lateral Casimir force can be greatly enhanced with respect to the case of a homogeneous SiC system (corresponding to the crossing of the dashed lines), when $\omega_{\rm T,p}>\omega_{\rm T,s}$, and consequently the particle and surface resonances overlap.

\begin{figure}
\begin{center}
\includegraphics[width=70mm,angle=0,clip]{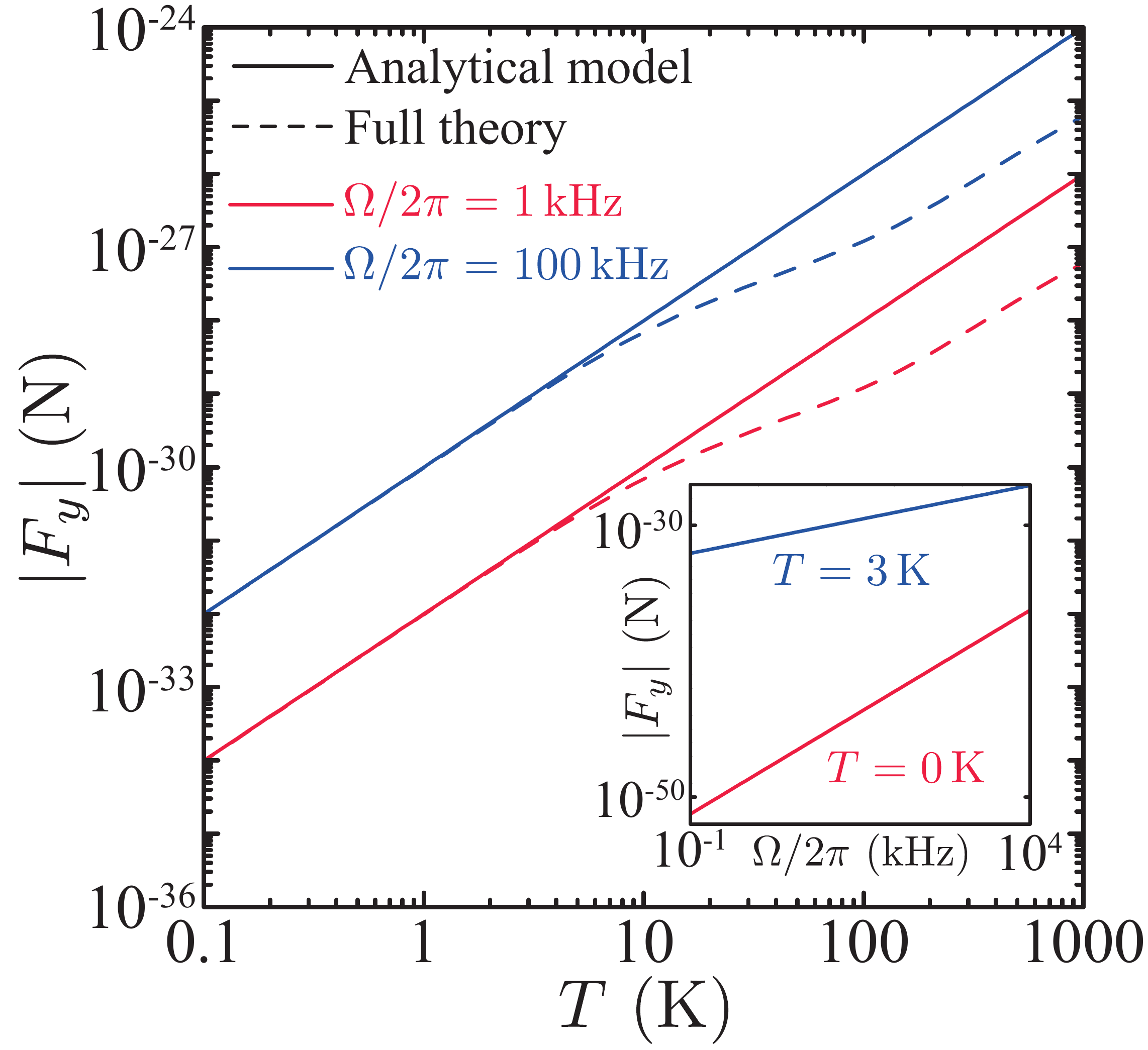}
\caption{Lateral force experienced by graphite spherical nanoparticle of radius $R=10\,$nm plotted as a function of the particle and surface temperatures, which are taken to be equal ($T_0=T_1=T$), for two different rotation frequencies. The dashed curves correspond to calculations using the full theory, while solid curves indicate the results obtained with the analytical model described in the text. The inset shows the lateral force as a function of the rotation frequency for two different temperatures calculated using the analytical model. In all cases the particle-surface distance is $d=3R$.} \label{fig5}
\end{center}
\end{figure}

{\it Analytical limit for metallic media.--}  It is possible to obtain a closed-form analytical expression of the lateral Casimir force for the case of metallic materials, whose response at low frequencies, well below interband transitions, can be described in terms of the static conductivity $\sigma_0$ using a Drude dielectric function $\varepsilon(\omega)=1+4\pi i\sigma_0/\omega$. When the relevant frequencies $\Omega$ and $k_{\rm B} T/\hbar$ are much smaller than $\sigma_0$, we can approximate ${\rm Im}\{\alpha\}\approx (3\omega R^3)/(4\pi\sigma_0)$, ${\rm Im}\{r_p\}\approx\omega/(2\pi\sigma_0)$, and ${\rm Im}\{\partial_yG_{yz}\}\approx (3\omega)/(32\pi d^4\sigma_0)$. Using these expressions into Eq.~(\ref{lateralforcef}) we obtain the following result for the lateral force
\begin{equation}\label{anmodel}
F_{y} =  -\frac{3\hbar}{256\pi^3\sigma_0^2}\frac{R^3}{d^4}\left[\frac{4\pi^2 k^2_{\rm B}}{\hbar^2}\left(T_1^2+T_0^2\right)+2\Omega^2\right]\Omega.
\end{equation} 
Interestingly, multiplying $F_y$ by $d$ we obtain the expression of the torque acting on the particle that was derived in Ref. \cite{ama19}, exactly as one would expect for a wheel of radius $d$ based on classical-mechanics arguments, which is consistent with the conservation of angular momentum. This hence reinforces our interpretation of the studied system as the Casimir analogue of a mechanical wheel rotating and moving on a surface, but without the necessity of contact between them.

%Figure~\ref{fig5} shows the lateral force for a $R=10\,$nm particle calculated using the analytical model (solid lines) and plotted as a function of the temperature $T$ (we take $T_1=T_0=T$) for two different $\Omega$. Both the particle and the surface are made of graphite, for which $\sigma_0=2.1\times10^{14}\,$s$^{-1}$, and their separation is $d=3R$. As expected, the results of the analytical model agree perfectly with the those obtained from the full theory (dashed lines) for low temperatures ($T\lesssim 10 \,$K), and they start to depart when the relevant frequencies approach to $\sigma_0$.

Figure~\ref{fig5} shows the lateral force for a $R=10\,$nm particle made of graphite, for which $\sigma_0=2.1\times10^{14}\,$s$^{-1}$, rotating at a distance $d=3R$ from a surface of the same material, plotted as a function of the temperature $T$ (we take $T_1=T_0=T$). The solid curves correspond to the results of the analytical model, while the dashed curves show the calculations obtained with the full theory using a tabulated dielectric function \cite{D03}. As expected, the analytical model agrees well with the full theory for low temperatures ($T\lesssim 10 \,$K), for which the lateral force shows a quadratic dependence on the temperature, as expected from Eq.~(\ref{anmodel}). 
Incidentally, Eq.~(\ref{anmodel}) also predicts a cubic dependence of the lateral force on the rotation frequency at zero temperature (\textit{i.e.}, $F_y\propto \Omega^3$ for $T_1=T_0=0\,$K), which is in sharp contrast with the linear behavior obtained for finite temperatures. These two behaviors are shown in the inset of Fig.~\ref{fig5} for rotation frequencies $\Omega/2\pi = 10^{-1} - 10^{4}\,$kHz.
  
{\it Concluding remarks.--} In summary, we have predicted the existence of a lateral Casimir force acting on rotating particles near planar surfaces. This force is enabled by the symmetry breaking induced by the particle rotation on the left- and right-handed components of the vacuum and thermal fluctuations of the particle dipole. This force is intimately related to the recently discovered lateral optical force acting on circularly polarized dipoles placed near a surface. 
The sign and magnitude of the lateral Casimir force depends on the geometrical and material properties of both the particle and the surface, allowing tunability of the force direction and magnitude and even complete suppression at certain particle-surface separations. The presented results describe a new type of lateral Casimir force acting on nanostructures, which is important for understanding, engineering, and controlling dynamic dispersion interactions at the nanoscale.

{\it Acknowledgements.--} A.M. acknowledges financial support from the Department of Physics and Astronomy and the College of Arts and Sciences of the University of New Mexico, and the UNM Center for Advanced Research Computing for computational resources used in this work. F.J.G.deA. acknowledges support from the Spanish MINECO (MAT2014-59096-P and SEV2015-0522), Fundaci\'o privada CELLEX, and AGAUR (2014-SGR- 1400). F.J.R.F. and A.V.Z. work was supported by EPSRC (UK) and the ERC iPLASMM project (321268). A.V.Z. acknowledges support from the Royal Society and the Wolfson Foundation.

\onecolumngrid

\appendix
\section{Appendix}
\subsection{Derivation of the lateral Casimir force}

The system under consideration is described in Fig.~\ref{fig1}(a) of the main text. It consists of a  nanoparticle, rotating with angular frequency $\Omega$ around the $x$-axis and placed a distance $d$ away from a planar surface, measured from its center.  We assume that the particle is small enough to be safely described within the dipolar approximation. In such limit, the lateral force exerted by an electric field $\EE$ on an electric dipole $\pp$ located at $\rr_0$ is given by
\begin{equation}
F_y=\sum_{i=x,y,z} p_i(t)\partial_y E_i(\rr_0,t). \nonumber
\end{equation}
The Casimir force originates in two different types of fluctuations: (i) fluctuations of the dipole moment of the particle, and (ii) fluctuations of the field created by random currents on the surface. Therefore, we can write
\begin{equation}\label{f1}
F_y=\sum_{i=x,y,z}\langle p_i^{\rm fl}(t)\partial_y E_i^{\rm ind}(\rr_0,t)+p_i^{\rm ind}(t)\partial_y E_i^{\rm fl}(\rr_0,t)\rangle,
\end{equation}
where $\langle\rangle$ denotes the average over fluctuations. It is important to notice that we do not need to include cross terms involving both dipole and field fluctuations, as these are uncorrelated because they originate in different physical systems. At this point it is convenient to work in the frequency domain $\omega$, defined via the Fourier transform
\begin{equation}
\EE(t)=\frac{1}{2\pi} \int_{-\infty}^{\infty}d\omega\,  \EE(\omega) e^{-i\omega t}\nonumber
\end{equation}
for the electric field, and similarly for other quantities. This allows us to write the induced field  in terms of the fluctuating dipole by using the Green function of the surface $\GG$
\begin{equation}
\EE^{\rm ind}(\rr,\omega)=\GG(\rr,\rr_0,\omega)\pp^{\rm fl}(\omega).\nonumber
\end{equation}
We only consider the Green function of the surface because the free-space component is not contributing to any Casimir force. In a similar way, the induced dipole can be expressed in terms of the fluctuating field with the help of the particle polarizability $\aa$
\begin{equation}
\pp^{\rm ind}(\omega)=\aa(\omega)\EE^{\rm fl}(\rr_0,\omega).\nonumber
\end{equation}
Here, we assume that $\aa$ is diagonal (\textit{i.e.}, $\alpha_{ij}=0$ for $i\neq j$).  
Using these expressions, the two terms of Eq.~(\ref{f1}) become quadratic in the dipole and field fluctuations. This allows us to compute the average over fluctuations using the fluctuation-dissipation theorem (FDT) \cite{N1928,CW1951} (see \cite{ama7} for a comprehensive derivation). The explicit expression of the FDT for dipole fluctuations is 
\begin{equation}\label{fdtp}
\langle p_i(\omega)p_j(\omega')\rangle=4\pi\hbar\delta(\omega+\omega') \mbox{Im}\left\{\alpha_{ij}(\omega)\right\}\left[n(T,\omega)+\frac{1}{2}\right],
\end{equation}
while for the field fluctuations we have
\begin{equation}\label{fdte}
\langle E_i(\rr,\omega)E_j(\rr', \omega')\rangle=4\pi\hbar\delta(\omega+\omega') \mbox{Im}\left\{G_{ij}(\rr,\rr',\omega)\right\}\left[n(T,\omega)+\frac{1}{2}\right].
\end{equation}
In these expressions, $n(T,\omega)= [\exp(\frac{\hbar\omega}{k_{ B} T}) - 1]^{-1} $ is the Bose-Einstein distribution function at temperature $T$.

Equipped with these tools, we can compute  the first term of Eq.~(\ref{f1}), which arises from the dipole fluctuations and can be written as
\begin{equation}\label{fp}
F_{y,\rm p}=\sum_{i=x,y,z}\langle p_i^{\rm fl}(t)\partial_y E_i^{\rm ind}(\rr_0,t)\rangle=\sum_{i,j=x,y,z}\int_{-\infty}^{\infty}\frac{d\omega d\omega'}{4\pi^2}e^{-i(\omega+\omega')t}\langle p_i^{\rm fl}(\omega) \partial_y G_{ij}(\rr_0,\rr_0,\omega')p_j^{\rm fl}(\omega') \rangle.
\end{equation}
Notice that the partial derivative only acts on the first coordinate of the Green function. In the following, for the sake of clarity, we do not show explicitly the spatial dependence of $\GG$.
In order to apply the FDT to average over dipole fluctuations, we need to write the dipoles in the frame rotating with the particle because it is only in that frame where the electronic and vibrational transitions that produce polarization are well defined. Taking into account that the particle rotates around the $x$-axis, the dipole components in the laboratory frame (unprimed) are expressed in terms of the ones of the rotating frame (primed) as
\begin{align}
p_x(\omega) & = p_{x'}(\omega), \nonumber \\
p_y(\omega) & = \frac1{2}\left[p_{y'}(\omega_+) + i p_{z'}(\omega_+) + p_{y'}(\omega_-) - i p_{z'}(\omega_-)\right], \nonumber \\
p_z(\omega) & = \frac1{2}\left[-ip_{y'}(\omega_+) + p_{z'}(\omega_+) + i p_{y'}(\omega_-) + p_{z'}(\omega_-)\right],  \nonumber
\end{align}
where $\omega_{\pm}=\omega\pm\Omega$. Inserting these expressions into Eq.~(\ref{fp}) and performing the average over fluctuations with the help of Eq.~(\ref{fdtp}), we obtain
\begin{align}
F_{y,\rm p}=&{}\frac{i\hbar}{4\pi}\int_{-\infty}^{\infty}d\omega \left[\partial_y G^{\ast}_{yz}(\omega)-\partial_yG^{\ast}_{zy}(\omega)\right] \Im\left\{\alpha_{yy}(\omega_+)+\alpha_{zz}(\omega_+)\right\}\left[n(T_1,\omega_+)+\frac{1}{2}\right] \nonumber \\
&-\frac{i\hbar}{4\pi}\int_{-\infty}^{\infty}d\omega \left[\partial_y G^{\ast}_{yz}(\omega)-\partial_yG^{\ast}_{zy}(\omega)\right]\Im\left\{\alpha_{yy}(\omega_-)+\alpha_{zz}(\omega_-)\right\}\left[n(T_1,\omega_-)+\frac{1}{2}\right],\nonumber
\end{align}
where $T_1$ is the particle temperature and $^{\ast}$ denotes the complex conjugate. In the derivation of this expression we use the fact that $\partial_y G_{xx}(\omega)=\partial_y G_{yy}(\omega)=\partial_y G_{zz}(\omega)=0$ for any translationally invariant surface. 
The integrals over frequency can be simplified by taking into account that $n(T,-\omega)=-n(T,\omega)-1$, and that, due to causality, $\alpha_{ij}(-\omega)=\alpha_{ij}^{\ast}(\omega)$ and $G_{ij}(-\omega)=G_{ij}^{\ast}(\omega)$. Therefore, making use of these symmetries, we find
\begin{align}\label{fp2}
F_{y,\rm p}=&{}\frac{\hbar}{2\pi}\int_{0}^{\infty}d\omega \Im\left\{\partial_y G_{yz}(\omega)-\partial_yG_{zy}(\omega)\right\} \Im\left\{\alpha_{yy}(\omega_+)+\alpha_{zz}(\omega_+)\right\}\left[n(T_1,\omega_+)+\frac{1}{2}\right] \nonumber \\
&-\frac{\hbar}{2\pi}\int_{0}^{\infty}d\omega \Im\left\{\partial_y G_{yz}(\omega)-\partial_yG_{zy}(\omega)\right\}\Im\left\{\alpha_{yy}(\omega_-)+\alpha_{zz}(\omega_-)\right\}\left[n(T_1,\omega_-)+\frac{1}{2}\right].
\end{align}

The second term of Eq.~(\ref{f1}), associated with field fluctuations, can be computed in a similar way:
\begin{equation}\label{fe}
F_{y,\rm E}=\sum_{i=x,y,z}\langle p_i^{\rm ind}(t)\partial_y E_i^{\rm fl}(\rr_0,t)\rangle=\sum_{i,j=x,y,z}\int_{-\infty}^{\infty}\frac{d\omega d\omega'}{4\pi^2}e^{-i(\omega+\omega')t}\langle \alpha_{{\rm eff},ij}(\omega)E_j^{\rm fl}(\rr_0,\omega) \partial_y E_i^{\rm fl}(\rr_0,\omega') \rangle,
\end{equation}
where $\aa_{\rm eff}(\omega)$ is the effective polarizability obtained when transforming the particle polarizability from the rotating to the lab frame. The components of $\aa_{\rm eff}(\omega)$ are related to those of $\aa(\omega)$ as
\begin{align}
\alpha_{{\rm eff},xx}(\omega) & = \alpha_{xx}(\omega), \nonumber \\
\alpha_{{\rm eff},yy}(\omega) &=\alpha_{{\rm eff},zz}(\omega)  = \frac1{4}\left[\alpha_{yy}(\omega_+) +\alpha_{zz}(\omega_+) + \alpha_{yy}(\omega_-) + \alpha_{zz}(\omega_-)\right], \nonumber \\
\alpha_{{\rm eff},yz}(\omega)&=-\alpha_{{\rm eff},zy}(\omega) =\frac{i}{4}\left[\alpha_{yy}(\omega_+) +\alpha_{zz}(\omega_+) - \alpha_{yy}(\omega_-) - \alpha_{zz}(\omega_-)\right].\nonumber
\end{align}
Substituting these expressions into Eq.~(\ref{fe}) and taking the average over the fluctuations of the field using Eq.~(\ref{fdte}), we obtain
\begin{align}
F_{y,\rm E}=&{}\frac{i\hbar}{4\pi}\int_{-\infty}^{\infty}d\omega \Im\left\{\partial_y G_{yz}(\omega)-\partial_yG_{zy}(\omega)\right\} \left[\alpha_{yy}(\omega_+)+\alpha_{zz}(\omega_+)-\alpha_{yy}(\omega_-)-\alpha_{zz}(\omega_-)\right]\left[n(T_0,\omega)+\frac{1}{2}\right], \nonumber 
\end{align}
where $T_0$ is the surface temperature. Once again, exploiting the symmetry of $\alpha_{ij}(\omega)$, $G_{ij}(\omega)$, and $n(T,\omega)$ as functions of $\omega$, we can simplify this result to
\begin{align}\label{fe2}
F_{y,\rm E}=&{}-\frac{\hbar}{2\pi}\int_{0}^{\infty}d\omega \Im\left\{\partial_y G_{yz}(\omega)-\partial_yG_{zy}(\omega)\right\} \Im\left\{\alpha_{yy}(\omega_+)+\alpha_{zz}(\omega_+)-\alpha_{yy}(\omega_-)-\alpha_{zz}(\omega_-)\right\}\left[n(T_0,\omega)+\frac{1}{2}\right]. 
\end{align}
Finally, combining Eqs.~(\ref{fp2}) and (\ref{fe2}), we obtain the expression of the lateral Casimir force
\begin{align} 
F_y=&{}F_{y,\rm p} + F_{y,\rm E},\nonumber \\ =&{}\frac{\hbar}{2\pi}\int_{0}^{\infty}d\omega \Im\left\{\partial_y G_{yz}(\omega)-\partial_yG_{zy}(\omega)\right\} \left[\Im\left\{\alpha_{yy}(\omega_+)+\alpha_{zz}(\omega_+)\right\}N(\omega_+)-\Im\left\{\alpha_{yy}(\omega_-)+\alpha_{zz}(\omega_-)\right\}N(\omega_-)\right],\nonumber
\end{align}
where $N(\omega_{\pm})=n(T_1,\omega_{\pm})-n(T_0,\omega)$. This equation reduces to the expression given in the main paper when the particle has axial symmetry (\textit{i.e.}, $\alpha_{yy}(\omega)=\alpha_{zz}(\omega)$).

\subsection{Derivation of the gradient of the Green function}

In order to calculate the gradient of the Green function we start by writing it using the Weyl identity \cite{NH06} as
\begin{equation}\label{g}
\GG(\rr,\rr',\omega)=\frac{i}{2\pi}\int\frac{dQ_x dQ_y}{k_z}e^{i Q_x (x-x')}e^{i Q_y (y-y')}e^{ik_z(z+z')} \left[r_p(\omega,Q)\overline{M}_p + r_s(\omega,Q)\overline{M}_s\right].
\end{equation}
Here, $Q_x$ and $Q_y$ are the components of the wave vector parallel to the surface, while $k_z$ is the component perpendicular to it, which satisfies $\sqrt{Q^2+k_z^2}=k$, where $Q^2=Q_x^2+Q_y^2$ and $k=\omega/c$. Furthermore, $r_p$ and $r_s$ are the Fresnel reflection coefficients for $p$- and $s$-polarized waves, and
\begin{equation}
\overline{M}_p
= \begin{pmatrix}
-k_z^2\frac{Q_x^2}{Q^2} & -k^2_z\frac{Q_x Q_y}{Q^2} & -Q_xk_z\\
-k^2_z\frac{Q_x Q_y}{Q^2} & -k^2_z\frac{Q_y^2}{Q^2} & -Q_yk_z\\
Q_x k_z & Q_y k_z &  Q^2
\end{pmatrix}, \ \ \ \ \ \ 
\overline{M}_s
= \begin{pmatrix}
k^2\frac{Q_y^2}{Q^2} & -k^2\frac{Q_x Q_y}{Q^2} & 0\\
-k^2\frac{Q_x Q_y}{Q^2} & k^2\frac{Q_x^2}{Q^2} & 0\\
0 & 0 &  0
\end{pmatrix}.\nonumber
\end{equation}
Using  Eq.~(\ref{g}), we can write $\partial_y \GG(\rr,\rr',\omega)$ as
\begin{equation}
\partial_y \GG(\rr,\rr',\omega)=-\frac{1}{2\pi}\int\frac{dQ_x dQ_y}{k_z}e^{i Q_x (x-x')}e^{i Q_y (y-y')}e^{ik_z(z+z')} Q_y \left[r_p(\omega,Q)\overline{M}_p + r_s(\omega,Q)\overline{M}_s\right],\nonumber
\end{equation}
which in the limit $\rr\rightarrow \rr_0$ and $\rr'\rightarrow \rr_0$ becomes
\begin{equation}
\partial_y \GG(\rr_0,\rr_0,\omega)=-\frac{1}{2\pi}\int\frac{dQ_x dQ_y}{k_z}e^{2ik_zd} Q_y \left[r_p(\omega,Q)\overline{M}_p + r_s(\omega,Q)\overline{M}_s\right],\nonumber
\end{equation}
where $d$ is the distance between the particle center and the surface. The integrals in this expression can be simplified by writing $Q_x$ and $Q_y$ in terms of $Q$ and the azimuthal angle $\phi$, chosen in such way that $Q_x=Q\cos\phi$ and $Q_y=Q\sin\phi$. 
By doing so, we can compute the integral over $\phi$ to finally obtain
\begin{equation}
\partial_y \GG(\rr_0,\rr_0,\omega)=\frac1{2}\int_0^{\infty}dQe^{2ik_zd} Q^3 r_p(\omega,Q)
\begin{pmatrix}
0 & 0 & 0\\
0 &0 &1\\
0 & -1 &  0
\end{pmatrix},\nonumber
\end{equation}
 which is the expression given in the main paper.

\twocolumngrid

%\bibliographystyle{apsrev}
%\bibliography{ref}

\begin{thebibliography}{39}
\expandafter\ifx\csname natexlab\endcsname\relax\def\natexlab#1{#1}\fi
\expandafter\ifx\csname bibnamefont\endcsname\relax
  \def\bibnamefont#1{#1}\fi
\expandafter\ifx\csname bibfnamefont\endcsname\relax
  \def\bibfnamefont#1{#1}\fi
\expandafter\ifx\csname citenamefont\endcsname\relax
  \def\citenamefont#1{#1}\fi
\expandafter\ifx\csname url\endcsname\relax
  \def\url#1{\texttt{#1}}\fi
\expandafter\ifx\csname urlprefix\endcsname\relax\def\urlprefix{URL }\fi
\providecommand{\bibinfo}[2]{#2}
\providecommand{\eprint}[2][]{\url{#2}}

\bibitem[{\citenamefont{Casimir}(1948)}]{C1948}
\bibinfo{author}{\bibfnamefont{H.~B.~G.} \bibnamefont{Casimir}},
  \bibinfo{journal}{Proc. Kon. Ned. Akad. Wetenschap}
  \textbf{\bibinfo{volume}{51}}, \bibinfo{pages}{793} (\bibinfo{year}{1948}).

\bibitem[{\citenamefont{Lamoreaux}(2007)}]{L07}
\bibinfo{author}{\bibfnamefont{S.}~\bibnamefont{Lamoreaux}},
  \bibinfo{journal}{Phys.\ Today} \textbf{\bibinfo{volume}{60}},
  \bibinfo{pages}{40} (\bibinfo{year}{2007}).

\bibitem[{\citenamefont{Dalvit et~al.}(2011)\citenamefont{Dalvit, Milonni,
  Roberts, and da~Rosa}}]{DMR11}
\bibinfo{author}{\bibfnamefont{D.}~\bibnamefont{Dalvit}},
  \bibinfo{author}{\bibfnamefont{P.}~\bibnamefont{Milonni}},
  \bibinfo{author}{\bibfnamefont{D.}~\bibnamefont{Roberts}}, \bibnamefont{and}
  \bibinfo{author}{\bibfnamefont{F.}~\bibnamefont{da~Rosa}},
  \emph{\bibinfo{title}{Casimir Physics}}, Lecture Notes in Physics
  (\bibinfo{publisher}{Springer Berlin Heidelberg}, \bibinfo{year}{2011}).

\bibitem[{\citenamefont{Munday and Capasso}(2010)}]{MC10}
\bibinfo{author}{\bibfnamefont{J.~N.} \bibnamefont{Munday}} \bibnamefont{and}
  \bibinfo{author}{\bibfnamefont{F.}~\bibnamefont{Capasso}},
  \bibinfo{journal}{Int.\ J.\ Mod.\ Phys.\ A} \textbf{\bibinfo{volume}{25}},
  \bibinfo{pages}{2252} (\bibinfo{year}{2010}).

\bibitem[{\citenamefont{Ball}(2007)}]{B07_2}
\bibinfo{author}{\bibfnamefont{P.}~\bibnamefont{Ball}},
  \bibinfo{journal}{Nature} \textbf{\bibinfo{volume}{447}},
  \bibinfo{pages}{772} (\bibinfo{year}{2007}).

\bibitem[{\citenamefont{Rodriguez et~al.}(2011)\citenamefont{Rodriguez,
  Capasso, and Johnson}}]{RCJ11}
\bibinfo{author}{\bibfnamefont{A.~W.} \bibnamefont{Rodriguez}},
  \bibinfo{author}{\bibfnamefont{F.}~\bibnamefont{Capasso}}, \bibnamefont{and}
  \bibinfo{author}{\bibfnamefont{S.~G.} \bibnamefont{Johnson}},
  \bibinfo{journal}{Nat.\ Photon.} \textbf{\bibinfo{volume}{5}},
  \bibinfo{pages}{211} (\bibinfo{year}{2011}).

\bibitem[{\citenamefont{Henkel et~al.}(2002)\citenamefont{Henkel, Joulain,
  Mulet, and Greffet}}]{HJM02}
\bibinfo{author}{\bibfnamefont{C.}~\bibnamefont{Henkel}},
  \bibinfo{author}{\bibfnamefont{K.}~\bibnamefont{Joulain}},
  \bibinfo{author}{\bibfnamefont{J.~P.} \bibnamefont{Mulet}}, \bibnamefont{and}
  \bibinfo{author}{\bibfnamefont{J.~J.} \bibnamefont{Greffet}},
  \bibinfo{journal}{J.\ Opt.\ A: Pure Appl.\ Opt.}
  \textbf{\bibinfo{volume}{4}}, \bibinfo{pages}{S109} (\bibinfo{year}{2002}).

\bibitem[{\citenamefont{Sushkov et~al.}(2011)\citenamefont{Sushkov, Kim,
  Dalvit, and Lamoreaux}}]{SKD11}
\bibinfo{author}{\bibfnamefont{A.~O.} \bibnamefont{Sushkov}},
  \bibinfo{author}{\bibfnamefont{W.~J.} \bibnamefont{Kim}},
  \bibinfo{author}{\bibfnamefont{D.~A.~R.} \bibnamefont{Dalvit}},
  \bibnamefont{and} \bibinfo{author}{\bibfnamefont{S.~K.}
  \bibnamefont{Lamoreaux}}, \bibinfo{journal}{Nat.\ Phys.}
  \textbf{\bibinfo{volume}{7}}, \bibinfo{pages}{230} (\bibinfo{year}{2011}).

\bibitem[{\citenamefont{Golestanian}(1997)}]{GK1997}
\bibinfo{author}{\bibfnamefont{R.~.~K.} \bibnamefont{Golestanian}},
  \bibinfo{journal}{Phys.\ Rev.\ Lett.} \textbf{\bibinfo{volume}{78}},
  \bibinfo{pages}{3421} (\bibinfo{year}{1997}).

\bibitem[{\citenamefont{Emig et~al.}(2001)\citenamefont{Emig, Hanke,
  Golestanian, and Kardar}}]{EHG01}
\bibinfo{author}{\bibfnamefont{T.}~\bibnamefont{Emig}},
  \bibinfo{author}{\bibfnamefont{A.}~\bibnamefont{Hanke}},
  \bibinfo{author}{\bibfnamefont{R.}~\bibnamefont{Golestanian}},
  \bibnamefont{and} \bibinfo{author}{\bibfnamefont{M.}~\bibnamefont{Kardar}},
  \bibinfo{journal}{Phys.\ Rev.\ Lett.} \textbf{\bibinfo{volume}{87}},
  \bibinfo{pages}{260402} (\bibinfo{year}{2001}).

\bibitem[{\citenamefont{Chen et~al.}(2002)\citenamefont{Chen, Mohideen,
  Klimchitskaya, and Mostepanenko}}]{CMK02_2}
\bibinfo{author}{\bibfnamefont{F.}~\bibnamefont{Chen}},
  \bibinfo{author}{\bibfnamefont{U.}~\bibnamefont{Mohideen}},
  \bibinfo{author}{\bibfnamefont{G.~L.} \bibnamefont{Klimchitskaya}},
  \bibnamefont{and} \bibinfo{author}{\bibfnamefont{V.~M.}
  \bibnamefont{Mostepanenko}}, \bibinfo{journal}{Phys.\ Rev.\ Lett.}
  \textbf{\bibinfo{volume}{88}}, \bibinfo{pages}{101801}
  (\bibinfo{year}{2002}).

\bibitem[{\citenamefont{Dalvit et~al.}(2008)\citenamefont{Dalvit, Neto,
  Lambrecht, and Reynaud}}]{DML08}
\bibinfo{author}{\bibfnamefont{D.~A.~R.} \bibnamefont{Dalvit}},
  \bibinfo{author}{\bibfnamefont{P.~A.~M.} \bibnamefont{Neto}},
  \bibinfo{author}{\bibfnamefont{A.}~\bibnamefont{Lambrecht}},
  \bibnamefont{and} \bibinfo{author}{\bibfnamefont{S.}~\bibnamefont{Reynaud}},
  \bibinfo{journal}{J.\ Phys.\ A: Math.\ Theor.} \textbf{\bibinfo{volume}{41}},
  \bibinfo{pages}{164028} (\bibinfo{year}{2008}).

\bibitem[{\citenamefont{Emig et~al.}(2003)\citenamefont{Emig, Hanke,
  Golestanian, and Kardar}}]{EHG03}
\bibinfo{author}{\bibfnamefont{T.}~\bibnamefont{Emig}},
  \bibinfo{author}{\bibfnamefont{A.}~\bibnamefont{Hanke}},
  \bibinfo{author}{\bibfnamefont{R.}~\bibnamefont{Golestanian}},
  \bibnamefont{and} \bibinfo{author}{\bibfnamefont{M.}~\bibnamefont{Kardar}},
  \bibinfo{journal}{Phys.\ Rev.\ A} \textbf{\bibinfo{volume}{67}},
  \bibinfo{pages}{022114} (\bibinfo{year}{2003}).

\bibitem[{\citenamefont{Blagov et~al.}(2004)\citenamefont{Blagov,
  Klimchitskaya, Mohideen, and Mostepanenko}}]{BKM04}
\bibinfo{author}{\bibfnamefont{E.~V.} \bibnamefont{Blagov}},
  \bibinfo{author}{\bibfnamefont{G.~L.} \bibnamefont{Klimchitskaya}},
  \bibinfo{author}{\bibfnamefont{U.}~\bibnamefont{Mohideen}}, \bibnamefont{and}
  \bibinfo{author}{\bibfnamefont{V.~M.} \bibnamefont{Mostepanenko}},
  \bibinfo{journal}{Phys.\ Rev.\ A} \textbf{\bibinfo{volume}{69}},
  \bibinfo{pages}{044103} (\bibinfo{year}{2004}).

\bibitem[{\citenamefont{Ashourvan et~al.}(2007)\citenamefont{Ashourvan, Miri,
  and Golestanian}}]{AMG07}
\bibinfo{author}{\bibfnamefont{A.}~\bibnamefont{Ashourvan}},
  \bibinfo{author}{\bibfnamefont{M.}~\bibnamefont{Miri}}, \bibnamefont{and}
  \bibinfo{author}{\bibfnamefont{R.}~\bibnamefont{Golestanian}},
  \bibinfo{journal}{Phys.\ Rev.\ Lett.} \textbf{\bibinfo{volume}{98}},
  \bibinfo{pages}{140801} (\bibinfo{year}{2007}).

\bibitem[{\citenamefont{Chiu et~al.}(2010)\citenamefont{Chiu, Klimchitskaya,
  Marachevsky, Mostepanenko, and Mohideen}}]{CKM10}
\bibinfo{author}{\bibfnamefont{H.-C.} \bibnamefont{Chiu}},
  \bibinfo{author}{\bibfnamefont{G.~L.} \bibnamefont{Klimchitskaya}},
  \bibinfo{author}{\bibfnamefont{V.~N.} \bibnamefont{Marachevsky}},
  \bibinfo{author}{\bibfnamefont{V.~M.} \bibnamefont{Mostepanenko}},
  \bibnamefont{and} \bibinfo{author}{\bibfnamefont{U.}~\bibnamefont{Mohideen}},
  \bibinfo{journal}{Phys.\ Rev.\ B} \textbf{\bibinfo{volume}{81}},
  \bibinfo{pages}{115417} (\bibinfo{year}{2010}).

\bibitem[{\citenamefont{Nasiri et~al.}(2010)\citenamefont{Nasiri, Moradian, and
  Miri}}]{NMM10}
\bibinfo{author}{\bibfnamefont{M.}~\bibnamefont{Nasiri}},
  \bibinfo{author}{\bibfnamefont{A.}~\bibnamefont{Moradian}}, \bibnamefont{and}
  \bibinfo{author}{\bibfnamefont{M.}~\bibnamefont{Miri}},
  \bibinfo{journal}{Phys.\ Rev.\ E} \textbf{\bibinfo{volume}{82}},
  \bibinfo{pages}{037101} (\bibinfo{year}{2010}).

\bibitem[{\citenamefont{Manjavacas and {Garc\'{\i}a de
  Abajo}}(2010{\natexlab{a}})}]{ama7}
\bibinfo{author}{\bibfnamefont{A.}~\bibnamefont{Manjavacas}} \bibnamefont{and}
  \bibinfo{author}{\bibfnamefont{F.~J.} \bibnamefont{{Garc\'{\i}a de Abajo}}},
  \bibinfo{journal}{Phys.\ Rev.\ Lett.} \textbf{\bibinfo{volume}{105}},
  \bibinfo{pages}{113601} (\bibinfo{year}{2010}{\natexlab{a}}).

\bibitem[{\citenamefont{Manjavacas and {Garc\'{\i}a de
  Abajo}}(2010{\natexlab{b}})}]{ama9}
\bibinfo{author}{\bibfnamefont{A.}~\bibnamefont{Manjavacas}} \bibnamefont{and}
  \bibinfo{author}{\bibfnamefont{F.~J.} \bibnamefont{{Garc\'{\i}a de Abajo}}},
  \bibinfo{journal}{Phys.\ Rev.\ A} \textbf{\bibinfo{volume}{82}},
  \bibinfo{pages}{063827} (\bibinfo{year}{2010}{\natexlab{b}}).

\bibitem[{\citenamefont{Zhao et~al.}(2012)\citenamefont{Zhao, Manjavacas,
  Garc\'{\i}a~de Abajo, and Pendry}}]{ama19}
\bibinfo{author}{\bibfnamefont{R.}~\bibnamefont{Zhao}},
  \bibinfo{author}{\bibfnamefont{A.}~\bibnamefont{Manjavacas}},
  \bibinfo{author}{\bibfnamefont{F.~J.} \bibnamefont{Garc\'{\i}a~de Abajo}},
  \bibnamefont{and} \bibinfo{author}{\bibfnamefont{J.~B.}
  \bibnamefont{Pendry}}, \bibinfo{journal}{Phys.\ Rev.\ Lett.}
  \textbf{\bibinfo{volume}{109}}, \bibinfo{pages}{123604}
  (\bibinfo{year}{2012}).

\bibitem[{\citenamefont{Pendry}(1997)}]{P97}
\bibinfo{author}{\bibfnamefont{J.~B.} \bibnamefont{Pendry}},
  \bibinfo{journal}{J.\ Phys.\ Condens.\ Matter} \textbf{\bibinfo{volume}{9}},
  \bibinfo{pages}{10301} (\bibinfo{year}{1997}).

\bibitem[{\citenamefont{Bliokh and Nori}(2012)}]{BN12}
\bibinfo{author}{\bibfnamefont{K.~Y.} \bibnamefont{Bliokh}} \bibnamefont{and}
  \bibinfo{author}{\bibfnamefont{F.}~\bibnamefont{Nori}},
  \bibinfo{journal}{PRA} \textbf{\bibinfo{volume}{85}}, \bibinfo{pages}{061801}
  (\bibinfo{year}{2012}).

\bibitem[{\citenamefont{Bliokh et~al.}(2014)\citenamefont{Bliokh, Bekshaev, and
  Nori}}]{BBN14}
\bibinfo{author}{\bibfnamefont{K.~Y.} \bibnamefont{Bliokh}},
  \bibinfo{author}{\bibfnamefont{A.~Y.} \bibnamefont{Bekshaev}},
  \bibnamefont{and} \bibinfo{author}{\bibfnamefont{F.}~\bibnamefont{Nori}},
  \bibinfo{journal}{Nat.\ Commun.} \textbf{\bibinfo{volume}{5}},
  \bibinfo{pages}{3300} (\bibinfo{year}{2014}).

\bibitem[{\citenamefont{Aiello et~al.}(2015)\citenamefont{Aiello, Banzer,
  Neugebauer, and Leuchs}}]{ABN15}
\bibinfo{author}{\bibfnamefont{A.}~\bibnamefont{Aiello}},
  \bibinfo{author}{\bibfnamefont{P.}~\bibnamefont{Banzer}},
  \bibinfo{author}{\bibfnamefont{M.}~\bibnamefont{Neugebauer}},
  \bibnamefont{and} \bibinfo{author}{\bibfnamefont{G.}~\bibnamefont{Leuchs}},
  \bibinfo{journal}{Nat.\ Photon.} \textbf{\bibinfo{volume}{9}},
  \bibinfo{pages}{789} (\bibinfo{year}{2015}).

\bibitem[{\citenamefont{Bliokh et~al.}(2015)\citenamefont{Bliokh,
  Rodr{\'\i}guez-Fortu{\~n}o, Nori, and Zayats}}]{BRN15}
\bibinfo{author}{\bibfnamefont{K.~Y.} \bibnamefont{Bliokh}},
  \bibinfo{author}{\bibfnamefont{F.~J.}
  \bibnamefont{Rodr{\'\i}guez-Fortu{\~n}o}},
  \bibinfo{author}{\bibfnamefont{F.}~\bibnamefont{Nori}}, \bibnamefont{and}
  \bibinfo{author}{\bibfnamefont{A.~V.} \bibnamefont{Zayats}},
  \bibinfo{journal}{Nat.\ Photon.} \textbf{\bibinfo{volume}{9}},
  \bibinfo{pages}{796} (\bibinfo{year}{2015}).

\bibitem[{\citenamefont{Scheel et~al.}(2015)\citenamefont{Scheel, Buhmann,
  Clausen, and Schneeweiss}}]{SBC15}
\bibinfo{author}{\bibfnamefont{S.}~\bibnamefont{Scheel}},
  \bibinfo{author}{\bibfnamefont{S.~Y.} \bibnamefont{Buhmann}},
  \bibinfo{author}{\bibfnamefont{C.}~\bibnamefont{Clausen}}, \bibnamefont{and}
  \bibinfo{author}{\bibfnamefont{P.}~\bibnamefont{Schneeweiss}},
  \bibinfo{journal}{PRA} \textbf{\bibinfo{volume}{92}}, \bibinfo{pages}{043819}
  (\bibinfo{year}{2015}).

\bibitem[{\citenamefont{Sukhov et~al.}(2015)\citenamefont{Sukhov,
  Kajorndejnukul, Naraghi, and Dogariu}}]{SKN15}
\bibinfo{author}{\bibfnamefont{S.}~\bibnamefont{Sukhov}},
  \bibinfo{author}{\bibfnamefont{V.}~\bibnamefont{Kajorndejnukul}},
  \bibinfo{author}{\bibfnamefont{R.~R.} \bibnamefont{Naraghi}},
  \bibnamefont{and} \bibinfo{author}{\bibfnamefont{A.}~\bibnamefont{Dogariu}},
  \bibinfo{journal}{Nat.\ Photon.} \textbf{\bibinfo{volume}{9}},
  \bibinfo{pages}{809} (\bibinfo{year}{2015}).

\bibitem[{\citenamefont{Rodr{\'\i}guez-Fortu{\~n}o
  et~al.}(2015)\citenamefont{Rodr{\'\i}guez-Fortu{\~n}o, Engheta,
  Mart{\'\i}nez, and Zayats}}]{REM15}
\bibinfo{author}{\bibfnamefont{F.~J.}
  \bibnamefont{Rodr{\'\i}guez-Fortu{\~n}o}},
  \bibinfo{author}{\bibfnamefont{N.}~\bibnamefont{Engheta}},
  \bibinfo{author}{\bibfnamefont{A.}~\bibnamefont{Mart{\'\i}nez}},
  \bibnamefont{and} \bibinfo{author}{\bibfnamefont{A.~V.}
  \bibnamefont{Zayats}}, \bibinfo{journal}{Nat.\ Commun.}
  \textbf{\bibinfo{volume}{6}}, \bibinfo{pages}{8799} (\bibinfo{year}{2015}).

\bibitem[{\citenamefont{Rodr{\'\i}guez-Fortu{\~n}o
  et~al.}(2013)\citenamefont{Rodr{\'\i}guez-Fortu{\~n}o, Marino, Ginzburg,
  O{\textquoteright}Connor, Mart{\'\i}nez, Wurtz, and Zayats}}]{RMG13}
\bibinfo{author}{\bibfnamefont{F.~J.}
  \bibnamefont{Rodr{\'\i}guez-Fortu{\~n}o}},
  \bibinfo{author}{\bibfnamefont{G.}~\bibnamefont{Marino}},
  \bibinfo{author}{\bibfnamefont{P.}~\bibnamefont{Ginzburg}},
  \bibinfo{author}{\bibfnamefont{D.}~\bibnamefont{O{\textquoteright}Connor}},
  \bibinfo{author}{\bibfnamefont{A.}~\bibnamefont{Mart{\'\i}nez}},
  \bibinfo{author}{\bibfnamefont{G.~A.} \bibnamefont{Wurtz}}, \bibnamefont{and}
  \bibinfo{author}{\bibfnamefont{A.~V.} \bibnamefont{Zayats}},
  \bibinfo{journal}{Science} \textbf{\bibinfo{volume}{340}},
  \bibinfo{pages}{328} (\bibinfo{year}{2013}).

\bibitem[{\citenamefont{Mueller and Capasso}(2013)}]{MC13}
\bibinfo{author}{\bibfnamefont{J.~P.~B.} \bibnamefont{Mueller}}
  \bibnamefont{and} \bibinfo{author}{\bibfnamefont{F.}~\bibnamefont{Capasso}},
  \bibinfo{journal}{PRB} \textbf{\bibinfo{volume}{85}}, \bibinfo{pages}{121410}
  (\bibinfo{year}{2013}).

\bibitem[{\citenamefont{Kien and Rauschenbeutel}(2014)}]{LR14}
\bibinfo{author}{\bibfnamefont{F.~L.} \bibnamefont{Kien}} \bibnamefont{and}
  \bibinfo{author}{\bibfnamefont{A.}~\bibnamefont{Rauschenbeutel}},
  \bibinfo{journal}{PRA} \textbf{\bibinfo{volume}{90}}, \bibinfo{pages}{023805}
  (\bibinfo{year}{2014}).

\bibitem[{\citenamefont{Nyquist}(1928)}]{N1928}
\bibinfo{author}{\bibfnamefont{H.}~\bibnamefont{Nyquist}},
  \bibinfo{journal}{Phys.\ Rev.} \textbf{\bibinfo{volume}{32}},
  \bibinfo{pages}{110} (\bibinfo{year}{1928}).

\bibitem[{\citenamefont{Callen and Welton}(1951)}]{CW1951}
\bibinfo{author}{\bibfnamefont{H.~B.} \bibnamefont{Callen}} \bibnamefont{and}
  \bibinfo{author}{\bibfnamefont{T.~A.} \bibnamefont{Welton}},
  \bibinfo{journal}{Phys.\ Rev.} \textbf{\bibinfo{volume}{83}},
  \bibinfo{pages}{34} (\bibinfo{year}{1951}).

\bibitem[{\citenamefont{Novotny and Hecht}(2006)}]{NH06}
\bibinfo{author}{\bibfnamefont{L.}~\bibnamefont{Novotny}} \bibnamefont{and}
  \bibinfo{author}{\bibfnamefont{B.}~\bibnamefont{Hecht}},
  \emph{\bibinfo{title}{Principles of Nano-Optics}}
  (\bibinfo{publisher}{Cambridge University Press}, \bibinfo{address}{New
  York}, \bibinfo{year}{2006}).

\bibitem[{\citenamefont{Gordon and Ashkin}(1980)}]{GA1980}
\bibinfo{author}{\bibfnamefont{J.~P.} \bibnamefont{Gordon}} \bibnamefont{and}
  \bibinfo{author}{\bibfnamefont{A.}~\bibnamefont{Ashkin}},
  \bibinfo{journal}{Phys.\ Rev.\ A} \textbf{\bibinfo{volume}{21}},
  \bibinfo{pages}{1606} (\bibinfo{year}{1980}).

\bibitem[{EPA()}]{EPAPS}
\bibinfo{note}{See supplementary material at
  http://link.aps.org/supplemental/xxx for more details on the theory.}

\bibitem[{\citenamefont{Palik}(1985)}]{P1985}
\bibinfo{author}{\bibfnamefont{E.~D.} \bibnamefont{Palik}},
  \emph{\bibinfo{title}{Handbook of Optical Constants of Solids}}
  (\bibinfo{publisher}{Academic Press}, \bibinfo{address}{San Diego},
  \bibinfo{year}{1985}).

\bibitem[{\citenamefont{Myroshnychenko
  et~al.}(2008)\citenamefont{Myroshnychenko, {Rodr\'{\i}guez-Fern\'andez},
  Pastoriza-Santos, Funston, Novo, Mulvaney, {Liz-Marz\'an}, and {Garc\'{\i}a
  de Abajo}}}]{paper112}
\bibinfo{author}{\bibfnamefont{V.}~\bibnamefont{Myroshnychenko}},
  \bibinfo{author}{\bibfnamefont{J.}~\bibnamefont{{Rodr\'{\i}guez-Fern\'andez}}},
  \bibinfo{author}{\bibfnamefont{I.}~\bibnamefont{Pastoriza-Santos}},
  \bibinfo{author}{\bibfnamefont{A.~M.} \bibnamefont{Funston}},
  \bibinfo{author}{\bibfnamefont{C.}~\bibnamefont{Novo}},
  \bibinfo{author}{\bibfnamefont{P.}~\bibnamefont{Mulvaney}},
  \bibinfo{author}{\bibfnamefont{L.~M.} \bibnamefont{{Liz-Marz\'an}}},
  \bibnamefont{and} \bibinfo{author}{\bibfnamefont{F.~J.}
  \bibnamefont{{Garc\'{\i}a de Abajo}}}, \bibinfo{journal}{Chem.\ Soc.\ Rev.}
  \textbf{\bibinfo{volume}{37}}, \bibinfo{pages}{1792} (\bibinfo{year}{2008}).

\bibitem[{\citenamefont{Draine}(2003)}]{D03}
\bibinfo{author}{\bibfnamefont{B.~T.} \bibnamefont{Draine}},
  \bibinfo{journal}{Astrophys.\ J.} \textbf{\bibinfo{volume}{598}},
  \bibinfo{pages}{1026} (\bibinfo{year}{2003}).

\end{thebibliography}

\end{document}